\documentclass[prb,showpacs,twocolumn,aps,superscriptaddress,a4paper]{revtex4}
\usepackage{dcolumn}
\usepackage{amsmath}
\usepackage{graphicx}
\usepackage{latexsym}  
\usepackage{amsfonts}  
\usepackage{amssymb} 
\usepackage{color}
\DeclareGraphicsExtensions{.eps,.pdf,.gif,.jpg,.png}  
  
\newcommand{\be}{\begin{equation}}  
\newcommand{\ee}{\end{equation}}
\newcommand{\beq}{\begin{eqnarray}}  
\newcommand{\eeq}{\end{eqnarray}}  
   
\newcommand{\hH}{\hat{H}}  
\newcommand{\hN}{\hat{N}} 
  
\begin{document}  
      
\def\bbe{\mbox{\boldmath $e$}}  
\def\bbf{\mbox{\boldmath $f$}}      
\def\bg{\mbox{\boldmath $g$}}  
\def\bh{\mbox{\boldmath $h$}}  
\def\bj{\mbox{\boldmath $j$}}  
\def\bq{\mbox{\boldmath $q$}}  
\def\bp{\mbox{\boldmath $p$}}  
\def\br{\mbox{\boldmath $r$}}      
  
\def\bone{\mbox{\boldmath $1$}}      
  
\def\dr{{\rm d}}  
  
\def\tb{\bar{t}}  
\def\zb{\bar{z}}  
  
\def\tgb{\bar{\tau}}

\def\bC{\mbox{\boldmath $C$}}  
\def\bG{\mbox{\boldmath $G$}}  
\def\bH{\mbox{\boldmath $H$}}  
\def\bK{\mbox{\boldmath $K$}}  
\def\bM{\mbox{\boldmath $M$}}  
\def\bN{\mbox{\boldmath $N$}}  
\def\bO{\mbox{\boldmath $O$}}  
\def\bQ{\mbox{\boldmath $Q$}}  
\def\bR{\mbox{\boldmath $R$}}  
\def\bS{\mbox{\boldmath $S$}}  
\def\bT{\mbox{\boldmath $T$}}  
\def\bU{\mbox{\boldmath $U$}}  
\def\bV{\mbox{\boldmath $V$}}  
\def\bZ{\mbox{\boldmath $Z$}}  
  
\def\bcalS{\mbox{\boldmath $\mathcal{S}$}}  
\def\bcalG{\mbox{\boldmath $\mathcal{G}$}}  
\def\bcalE{\mbox{\boldmath $\mathcal{E}$}}  
  
\def\bgG{\mbox{\boldmath $\Gamma$}}  
\def\bgL{\mbox{\boldmath $\Lambda$}}  
\def\bgS{\mbox{\boldmath $\Sigma$}}  
  
\def\bgr{\mbox{\boldmath $\rho$}}  
  
\def\a{\alpha}  
\def\b{\beta}  
\def\g{\gamma}  
\def\G{\Gamma}  
\def\d{\delta}  
\def\D{\Delta}  
\def\e{\epsilon}  
\def\ve{\varepsilon}  
\def\z{\zeta}  
\def\h{\eta}  
\def\th{\theta}  
\def\k{\kappa}  
\def\l{\lambda}  
\def\L{\Lambda}  
\def\m{\mu}  
\def\n{\nu}  
\def\x{\xi}  
\def\X{\Xi}  
\def\p{\pi}  
\def\P{\Pi}  
\def\r{\rho}  
\def\s{\sigma}  
\def\S{\Sigma}  
\def\t{\tau}  
\def\f{\phi}  
\def\vf{\varphi}  
\def\F{\Phi}  
\def\c{\chi}  
\def\w{\omega}  
\def\W{\Omega}  
\def\Q{\Psi}  
\def\q{\psi}  
  
\def\ua{\uparrow}  
\def\da{\downarrow}  
\def\de{\partial}  
\def\inf{\infty}  
\def\ra{\rightarrow}  
\def\bra{\langle}  
\def\ket{\rangle}  
\def\grad{\mbox{\boldmath $\nabla$}}  
\def\Tr{{\rm Tr}}  
\def\Re{{\rm Re}}  
\def\Im{{\rm Im}}

\title{Kadanoff-Baym approach to quantum transport through interacting nanoscale systems: From the transient to the steady-state regime}

\author{Petri My\"oh\"anen}
\affiliation{Department of Physics, Nanoscience Center, FIN 40014, University of Jyv\"askyl\"a,
Jyv\"askyl\"a, Finland}
\author{Adrian Stan}
\affiliation{Department of Physics, Nanoscience Center, FIN 40014, University of Jyv\"askyl\"a,
Jyv\"askyl\"a, Finland}
\author{Gianluca Stefanucci}
\affiliation{Dipartimento di Fisica, Universit\`a di Roma Tor Vergata, Via della Ricerca Scientifica 1, I-00133 Rome, Italy}
\affiliation{European Theoretical Spectroscopy Facility (ETSF)}
\author{Robert van Leeuwen}
\affiliation{Department of Physics, Nanoscience Center, FIN 40014, University of Jyv\"askyl\"a,
Jyv\"askyl\"a, Finland}
\affiliation{European Theoretical Spectroscopy Facility (ETSF)}

\date{\today}  

\begin{abstract}
We propose a time-dependent many-body approach to study the short-time dynamics 
of correlated electrons in quantum transport
through nanoscale systems contacted to metallic leads.
This approach is based on the time-propagation of the Kadanoff-Baym equations 
for the nonequilibrium many-body Green's function of
open and interacting systems out of equilibrium. 
An important feature of the method is that it takes full account of electronic correlations and embedding effects
in the presence of time-dependent external fields, while at the same time satisfying the charge conservation law.  
The method further extends the Meir-Wingreen formula to the time domain for initially correlated
states. We study the electron dynamics of a correlated quantum wire attached to two-dimensional leads
exposed to a sudden switch-on of a bias voltage using conserving many-body approximations at Hartree-Fock, second Born and GW level. 
We obtain detailed results for the transient currents,  dipole moments, spectral functions,
charging times, and the many-body screening of the quantum wire as 
well as for the time-dependent density pattern in the leads,
and we show how the time-dependence of these observables provides 
a wealth of information on the level structure of the quantum wire out of equilibrium.
For moderate interaction strenghts the 2B and GW results are in 
excellent agreement {\em at all times}.  
We find that many-body effects beyond the Hartree-Fock approximation 
have a large effect on the qualitative behavior of the system and
lead to a bias dependent gap closing and quasiparticle broadening, 
shortening of the transient times and washing out of the step features 
in the current-voltage curves.

\end{abstract}
  
  
\maketitle  

\section{Introduction}
The description of electron transport through nanoscale systems 
contacted to metallic leads is 
currently under intensive investigation especially due to 
the possibility of miniaturizing integrated devices
in electrical circuits.\cite{book} Several theoretical methods 
have been proposed to address the steady state properties of these 
systems. 

Ab initio formulations based on Time-Dependent (TD) Density Functional 
Theory\cite{rg.1984,vl.1999,sa.2004,sa2.2004,dvt.2004,ewk.2004} (DFT) and Current Density Functional 
Theory\cite{vk.1996,v.2004,szvdv.2005,kbe.2006,jbg.2007} provide a 
virtual exact framework to account for correlation effects both in 
the leads and the device but lack of a systematic route to improve the level of 
the approximations. {\em Ad hoc} approximations have been 
successfully implemented to describe qualitative features of the 
I/V characteristic of molecular junctions in the Coulomb blockade 
regime.\cite{tfsb.2005,p.2005,se.2008,sgc.2008} More sophisticated 
approximations are, however, needed for, e.g.,
non-resonant tunneling transport through weakly coupled 
molecules.\cite{ewk.2004,dvpl.2000,stbmo.2003,qvclhn.2007,kby.2007}

The possibility of including relevant physical processes through an  
insightful selection of Feynman diagrams is the main advantage of Many-Body 
Perturbation Theory (MBPT) over one-particle schemes. Even though 
computationally more expensive MBPT offers an invaluable tool to 
quantify the effects of electron correlations by analyzing, e.g.,  
the quasi-particle spectra, life-times, screened interactions, etc.
One of the most remarkable advances in the MBPT formulation of electron transport  
was given by Meir and Wingreen who provided an equation
for the steady state current through a correlated device region\cite{mw.1992,jwm.1994} 
thus generalizing the Landauer formula.\cite{l.1957} The 
Meir-Wingreen formula is cast in terms of the interacting Green's 
function and self-energy in the device region and can be approximated using standard 
diagrammatic techniques. Exploiting Wick's theorem\cite{FetterWalecka} a general diagram for 
the self-energy can be written in terms of bare Green's functions and 
interaction lines. Any approximation to the self-energy which 
contains a finite number of such diagrams does, however, violate many conservation
laws. Conserving approximations\cite{bk.1961,b.1962,vbdvls.2005,bbvl.2007} require the 
resummation of an infinite number of diagrams and 
are of paramount importance in nonequilibrium problems as they guarantee 
satisfaction of fundamental conservation laws such as charge conservation. 
Examples of conserving approximations are the Hartree-Fock (HF), 
second Born (2B), GW, T-matrix, and fluctuation exchange (FLEX) approximations.\cite{bs.1989,vlds.2006} 
The success of the GW approximation\cite{h.1965,ag.1998} in 
describing spectral features of atoms and 
molecules\cite{nrdsdcf.2005,sdvl.2006,sdvl.2009} as 
well as of interacting model clusters\cite{vgh.1995} prompted 
efforts to implement the Meir-Wingreen formula at the GW level
in simple molecular junctions and tight-binding 
models.\cite{tr.2007,dfmo.2007,wshm.2008,tr.2008,t.2008,shlm.2009}

The advantage of using molecular devices in future nanoelectronics
is, however, not only the miniaturization of integrated circuits. Nanodevices 
can work at the THertz regime and hence perform operations in few 
picoseconds or even faster. Space and time can both be considerably reduced. 
Nevertheless, at the sub-picosecond time scale stationary steady-state 
approaches are inadequate to extract crucial quantities like, e.g., the 
switching- or charging-time of a molecular diode, and 
consequently to understand how to optimize the device performance.
Despite the importance that an increase in the operational speed may 
have in practical applications, the ultrafast dynamical response 
of nanoscale devices is still largely unexplored.
This paper wants to make a further step towards the theoretical 
modeling of correlated TD quantum transport. 

Recently several practical schemes have been proposed to tackle 
TD quantum transport problems of noninteracting 
electrons.\cite{ksarg.2005,zmjg.2005,hhlkch.2006,mgm.2007,bcg.2008} 
In some of these schemes the electron-electron interaction can be included 
within a TDDFT framework\cite{sa.2004,ksarg.2005} and few calculations on 
the transient electron dynamics of molecular junctions have been performed at 
the level of the adiabatic local density 
approximation.\cite{cevt.2006,bsdv.2005,zwymc.2007} 
Alternatively, approaches based on Bohm 
trajectories\cite{o.2007,aso.2009} or on the density matrix renormalization 
group\cite{ahfrbd.2006} have been put forward to calculate 
TD currents and densities through interacting quantum systems.
So far, however, no one has extended the diagrammatic MBPT 
formulation of Meir and Wingreen to the time-domain. As in the 
steady-state case the MBPT formulation allows for including relevant 
scattering mechanisms via a proper selection of physically meaningful 
Feynman diagrams. The appealing nature of diagrammatic expansions 
renders MBPT an attractive alternative to investigate 
out-of-equilibrium systems.

In a recent Letter\cite{mssvl.2008} we proposed a time-dependent MBPT formulation of 
quantum transport which is based on the real-time propagation
of the Kadanoff-Baym (KB) equations\cite{dsvl.2006,dvl.2007,vfva.2009,bbvlds.2009,book2,d.1984,kb.2000}
for open and interacting systems. The KB equations are equations of motion for the
nonequilibrium Green's function from which basic properties
of the system can be calculated. It is the purpose of this paper to 
give a detailed account of the theoretical derivation and to extend 
the numerical analysis to quantum wires connected to two-dimensional 
leads. For practical calculations we have implemented the fully 
self-consistent HF, 2B and GW conserving approximations. Our results 
reduce to those of steady-state MBPT implementations in the long time 
limit. Having full access to the transient dynamics we are able, 
however, to extract novel information like the switching- and 
charging-times, the time-dependent renormalization of the electronic 
levels, the role of initial correlations, the time-dependent 
dipole moments etc. Furthermore, the non-locality in time of the 2B 
and GW self-energies allows us to highlight non-trivial memory 
effects occuring before the steady-state is reached. 
We also wish to emphasize that our approach is not limited to DC 
biases. Arbitrary driving fields like AC biases, voltage pulses, 
pumping fields, etc. can be dealt with at the same computational cost.

The paper is organized as follows. All derivations and formulas are 
given in Section \ref{theory}. We 
present the class of many-body systems that can be studied within our 
KB formulation in Section \ref{modelham} and derive the equations of motion for the 
nonequilibrium Green's function in the device region in Section 
\ref{keldeom} (see also Appendix \ref{realtimekb}).
The equations of motion are then used to prove the continuity 
equation for all conserving approximations, Section \ref{chcons}, 
and to extend the Meir-Wingreen formula to the time domain for initially 
correlated systems, Section \ref{emwformula}. Using an {\em inbedding} technique 
in Section \ref{denslead} we derive the main 
equations to calculate the time-dependent density in the leads.
In Section \ref{numres} we present the results of our TD
simulations for a one-dimensional wire connected to two-dimensional 
leads. The Keldysh Green's function, which is the basic quantity of the KB 
approach, of the open wire is studied in Section \ref{kgreensect} 
showing different time-dependent regimes relevant to the subsequent analysis.
In Section \ref{tdcurranddm} and \ref{tddipmom} we calculate the 
TD current and dipole moment respectively. We find that the 2B and GW 
results are in excellent agreement {\em at all times} 
and can differ substantially from the HF results. We also perform the 
Fourier analysis of the transient oscillations and reveal the 
underlying out-of-equilibrium electronic structure of the open 
wire.\cite{skrg.2008} The dynamically screened interaction of the GW approximation is 
investigated in Section \ref{tdw} with emphasis on the time-scales
of retardation effects. Section \ref{tdfriedel} is devoted to the study 
of the TD rearrangement of the density in the two-dimensional leads 
after the switch-on of an external bias. Such an analysis permits us to 
test the validity of a commonly used assumption in quantum transport, 
i.e., that the leads remain in thermal equilibrium. Finally, in 
Section \ref{conc} we draw our main conclusions and future 
perspectives.

\section{Theory}
\label{theory}

\subsection{The model Hamiltonian}
\label{modelham}
\begin{figure}[t]
\centering
\includegraphics[width=0.48\textwidth]{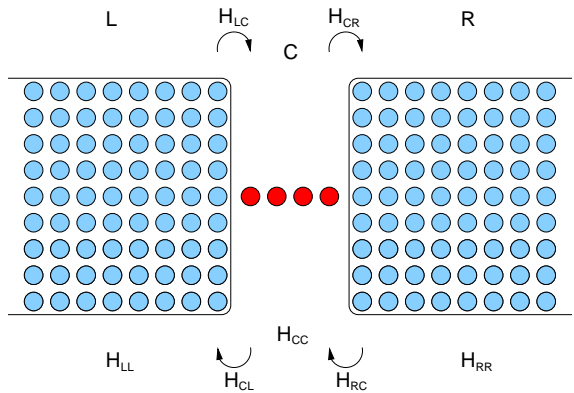}
\caption[]{Sketch of the transport setup. 
The correlated central region (C) is coupled to semi-infinite left (L) and right (R) tight-binding leads via tunneling
Hamiltonians $\mathbf{H}_{\alpha\textnormal{\scriptsize{C}}}$ 
and $\mathbf{H}_{\textnormal{\scriptsize{C}}\alpha}$, $\alpha={\rm L,R}$. }
\label{fig1}
\end{figure}

We consider a class of quantum correlated open systems (which we call 
central regions) coupled to noninteracting reservoirs (which we call 
leads), see Fig. \ref{fig1}.
The Hamiltonian has the general form
\be
\hH (t) = \hH_{\rm C} (t) + \sum_{\alpha} \hH_\alpha (t) + \hH_{T} - 
\mu \hN,
\label{eq:ham}
\ee
where $\hH_{\rm C}$, $\hH_{\a}$, $\hH_{T}$ are the central region, the lead $\alpha$ and the 
tunneling Hamiltonians respectively
and $\hN$ is the particle number operator coupled to chemical potential $\mu$.
We assume that there is no direct coupling between the leads.
The correlated central region has a Hamiltonian of the form
\be
\hH_{\rm C} (t)=
\sum_{ij,\sigma} h_{ij} (t) \hat{d}_{i\sigma}^{\dagger}\hat{d}_{j\sigma} +
\frac{1}{2}\sum_{\substack{ijkl\\ \sigma\sigma'}} 
v_{ijkl}\hat{d}_{i\sigma}^{\dagger}\hat{d}_{j\sigma'}^{\dagger}\hat{d}_{k\sigma'}\hat{d}_{l\sigma}, 
\label{eq:twobody}
\ee
where $i,j$ label a complete set of one-particle states in the central region, 
$\sigma,\sigma'$ are spin-indices and 
$\hat{d}^{\dagger},\hat{d}$ are the creation and annihilation 
operators respectively. 
The one-body part of the Hamiltonian $h_{ij}(t)$ may have an arbitrary time-dependence, 
describing, {\em e.g.}, a gate voltage
or pumping fields. The two-body part accounts for interactions between the 
electrons where $v_{ijkl}$ are, for example in the
case of a molecule, the standard two-electron integrals of the Coulomb interaction.
The lead Hamiltonians have the form
\be
\hH_\alpha (t) = U_{\a}(t) \hN_\alpha + 
\sum_{ij,\sigma} h_{ij}^{\alpha}  \, 
\hat{c}_{i\sigma\alpha}^{\dagger}\hat{c}_{j\sigma \alpha} ,
\label{eq:hamlead}
\ee
where the creation and annihilation operators for the leads are denoted by 
$\hat{c}^{\dagger}$ and $\hat{c}$.
Here $\hN_\alpha = \sum_{i,\sigma} \hat{c}_{i\sigma\alpha}^{\dagger}\hat{c}_{i\sigma \alpha}$ 
is the operator describing the
number of particles in lead $\alpha$.
The one-body part of the Hamiltonian $h_{ij}^\alpha$ describes 
metallic leads and can be calculated using a tight-binding representation, or 
a real-space grid or any other convenient basis set. We are interested 
in exposing the leads to an external electric field which varies on 
a time-scale much longer than the typical plasmon time-scale. 
Then, the coarse-grained time evolution can be performed assuming 
a perfect instantaneous screening in the leads 
and the homogeneous time-dependent field $U_\alpha (t)$ can be interpreted as the sum of 
the external and the screening field, {\em i.e.},
the applied bias. This effectively means that the leads are treated at a Hartree mean field level.
We finally consider the tunneling Hamiltonian $\hH_T$
\be
\hH_T  = \sum_{ij,\sigma \alpha} V_{i,j\alpha} [ \hat{d}_{i\sigma}^{\dagger}\hat{c}_{j\sigma \alpha} + 
\hat{c}_{j\sigma \alpha}^{\dagger}\hat{d}_{i\sigma} ]
\ee
which describes the coupling of the leads to the interacting central region. This completes 
the full description of the Hamiltonian of the system. In the next section we study the equations of motion for
the corresponding  Green's function.

\subsection{Equation of motion for the Keldysh Green's function}
\label{keldeom}

\begin{figure}[t]
\centering
\includegraphics[width=0.48\textwidth]{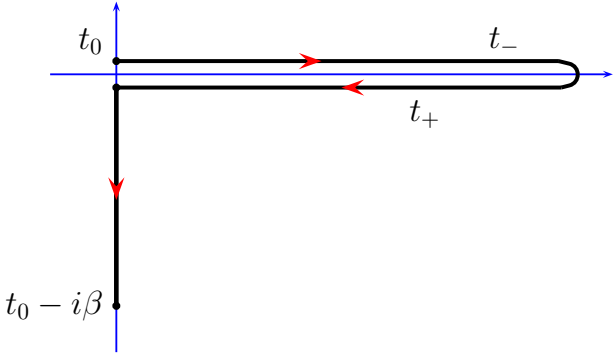}
\caption[]{Keldysh contour $\gamma$. Times on the upper/lower branch 
are specified with the subscript $\mp$.}
\label{fig2}
\end{figure}

We assume the system to be contacted and in equilibrium at inverse temperature $\beta$ 
before time $t=t_0$ and described by Hamiltonian $\hH_0$. For times $t > t_0$ the system is
driven out of equilibrium by an external bias and we aim to study the 
time-evolution of the electron density, current, etc..
In order to describe the electron dynamics in this system we use Keldysh Green's function theory
(for a review see Ref.\onlinecite{d.1984})
which allows us to include many-body effects in a diagrammatic way.
The Keldysh Green's function is
defined as the expectation value of the contour-ordered product 
\beq
\bcalG_{rs}(z,z') &=& -i \frac{\Tr \left\{ 
\mathcal{T} [e^{-i\int d\bar{z}\hH(\bar{z})}\hat{a}_r(z)\hat{a}^{\dagger}_s(z')] \right\}}
{\Tr \left\{ e^{-\beta \hH_0}\right\}} \nonumber \\
&=&
-i\langle \mathcal{T} [\hat{a}_r(z)\hat{a}^{\dagger}_s(z')] \rangle,
\label{kgreen}
\eeq
where $\hat{a}$ and $\hat{a}^\dagger$ are either lead or central region operators and the indices $r$ and $s$ are
collective indices for position and spin. The variable $z$ is a time contour variable that specifies the location of the operators
on the time contour.  The operator $\mathcal{T}$ orders the operators along the Keldysh contour
displayed in Fig. \ref{fig2}, consisting of two real time branches 
and the imaginary track running from $t_0$ to $t_0-i\beta$.
In the definition of the Green's function the trace is taken with respect to the many-body states of the system.\\
All time-dependent one-particle properties can be calculated from $\bcalG$. For instance,
the time-dependent density matrix is given as
\be
n_{rs} (t) = -i \bcalG_{rs} (t_{-},t_{+}),
\label{tddm}
\ee
where the times $t_{\pm}$ lie on the lower/upper branch of the contour.
The equations of motion for the Green's function of the full system can 
be easily derived from the definition Eq. (\ref{kgreen}) and read
\beq
 i\de_z  \bcalG(z,z') 
&=& \delta(z,z')\mathbf{1} + \mathbf{H}(z) \bcalG(z,z')  \nonumber \\
&+& \int d\bar{z}\,\bgS^{\rm MB}(z,\bar{z})\bcalG(\bar{z},z'),
\label{eqofmotion}\\
-i\de_{z'}\bcalG(z,z') &=& 
 \delta(z,z')\mathbf{1} + \bcalG (z,z') \mathbf{H} (z')\nonumber \\
&+& \int d\bar{z}\, \bcalG (z,\bar{z}) \bgS^{\rm{MB}}(\bar{z},z),
\label{eqofmotion_adj}
\eeq
where $\bgS^{\rm MB}$ 
is the many-body self-energy, $\mathbf{H}(z)$ is the matrix representation of 
the one-body part of the full Hamiltonian and the integration is performed over the Keldysh-contour. 
This equation of motion needs to be solved with the boundary 
conditions\cite{k.1957,ms.1959}
\begin{equation}
\begin{split}
\mbox{\boldmath $\mathcal{G}$}(t_0,z')&=-\mbox{\boldmath 
$\mathcal{G}$}(t_0-i\beta,z'),\\
\mbox{\boldmath $\mathcal{G}$}(z,t_0)&=-\mbox{\boldmath 
$\mathcal{G}$}(z,t_0-i\beta),
\end{split}
\label{kms}
\end{equation}
which follow directly from the definition of the Green's function 
Eq. (\ref{kgreen}).
Explicitly, the one-body Hamiltonian $\mathbf{H}$ 
for the case of two leads, Left (L) and Right (R) connected to a central region (C), is
\begin{equation}\label{1bham}
\mathbf{H} = 
\left[
\begin{array}{ccc}
\mathbf{H}_{\textnormal{\scriptsize{LL}}} & 
\mathbf{H}_{\textnormal{\scriptsize{LC}}} &  \mathbf{0} \\
\mathbf{H}_{\textnormal{\scriptsize{CL}}} & 
\mathbf{H}_{\textnormal{\scriptsize{CC}}} & 
\mathbf{H}_{\textnormal{\scriptsize{CR}}} \\
\mathbf{0} & \mathbf{H}_{\textnormal{\scriptsize{RC}}} & 
\mathbf{H}_{\textnormal{\scriptsize{RR}}}
\end{array}
\right]
\end{equation}
where the different block matrices describe the projections of the 
one-body part $\mathbf{H}$ of the Hamiltonian onto different 
subregions. They are explicitly given as
\beq
\left( \mathbf{H}_{\alpha \alpha} \right)_{i\sigma,j\sigma'}(z) &=& 
\left[ h_{ij}^\alpha + \delta_{ij} (U_\alpha (z) - \mu) \right] 
\delta_{\sigma \sigma'}, \\
\left( \mathbf{H}_{\rm{CC}} \right)_{i\sigma,j\sigma'}(z) &=& 
\left[ h_{ij}(z) - \delta_{ij} \mu \right] \delta_{\sigma \sigma'},\\
\left( \mathbf{H}_{\rm{C} \alpha} \right)_{i\sigma,j\sigma'} &=&  \left( \mathbf{H}_{\alpha \rm{C}}^\dagger \right)_{j\sigma',i\sigma,}= 
V_{i,j\alpha} \delta_{\sigma \sigma'} .
\label{HalphaC}
\eeq
We focus on the dynamical processes
occuring in the central region. These are described by the Green's 
function $\bcalG_{\rm{CC}}$ projected onto region C. 
We therefore want to extract from the block matrix structure for the Green's function
\begin{equation}
\mbox{\boldmath $\mathcal{G}$} = 
\left[
\begin{array}{ccc}
\bcalG_{\textnormal{\scriptsize{LL}}} &\bcalG_{\textnormal{\scriptsize{LC}}} & \bcalG_{\textnormal{\scriptsize{LR}}}\\
\bcalG_{\textnormal{\scriptsize{CL}}} & \bcalG_{\textnormal{\scriptsize{CC}}} & \bcalG_{\textnormal{\scriptsize{CR}}} \\
\bcalG_{\textnormal{\scriptsize{RL}}} &\bcalG_{\textnormal{\scriptsize{RC}}} & \bcalG_{\textnormal{\scriptsize{RR}}}
\end{array}
\right]
\end{equation}
an equation for $\bcalG_{\rm{CC}}$.
The many-body self-energy in Eq. (\ref{eqofmotion}) has nonvanishing entries only for indices in region C.
This is an immediate consequence of the fact that the diagrammatic expansion of the self-energy starts and ends with and
interaction line which in our case is confined in the central region (see last term of Eq. (\ref{eq:twobody})).
This also implies that $\bgS^{\rm{MB}} [\bcalG_{\rm{CC}}]$ is a functional of $\bcalG_{\rm{CC}}$ only.
From these considerations it follows that in the one-particle basis the matrix structure of $\bgS^{\rm{MB}}$ is given as
\begin{equation}
 \bgS^{\rm{MB}} = 
\left[
\begin{array}{ccc}
0 & 0 & 0\\
0 & \bgS^{\rm{MB}}_{\rm{CC}} [\bcalG_{\rm{CC}}] & 0 \\
0 & 0 & 0
\end{array}
\right].
\end{equation}
The projection of the equation of motion (\ref{eqofmotion}) onto 
regions ${\rm CC}$ and $\alpha {\rm C}$ yields
\begin{equation}\label{eomCC}
\begin{split}
&\Bigl\{ i\de_z\mathbf{1} - \mathbf{H}_{\rm CC}(z) \Bigr\}\bcalG_{\rm CC}(z,z') = 
\delta(z,z')\mathbf{1}\,\, + \\
&\sum_{\alpha}\mathbf{H}_{{\rm C}\alpha}\bcalG_{\alpha\rm C}(z,z') 
+ \int d\bar{z}\,\bgS_{\rm CC}^{\rm MB}(z,\bar{z})\bcalG_{\rm CC}(\bar{z},z')
\end{split}
\end{equation}
for the central region and
\be
\Bigl\{ i\de_z\mathbf{1} - \mathbf{H}_{\alpha \alpha}(z) \Bigr\} \bcalG_{\alpha \rm{C}} (z,z') = 
 \mathbf{H}_{\alpha \rm{C}}\bcalG_{\rm{CC}}(z,z') 
\label{eomCa}
\ee
for the projection on $\alpha {\rm C}$.
The latter equation can be solved for $\bcalG_{\alpha \rm{C}}$, 
taking into account the boundary conditions of Eq. (\ref{kms}), to yield
\begin{equation}
\begin{split}
\bcalG_{\alpha\rm C}(z,z') =  
\int d\bar{z}\,\mathbf{g}_{\alpha\alpha}(z,\bar{z})\,
\mathbf{H}_{\alpha\rm C}\bcalG_{\rm CC}(\bar{z},z'),
\end{split}
\label{GalphaC}
\end{equation}
where the integral is along the Keldysh contour.
Here we defined $\bg_{\alpha \alpha}$ as the solution of
\be
\Bigl\{i\partial_z\mathbf{1} - \mathbf{H}_{\alpha \alpha}(z)\Bigr\} \bg_{\alpha \alpha}(z,z') 
=\delta(z,z')\mathbf{1},
\ee
with boundary conditions Eq. (\ref{kms}). 
The function $\bg_{\alpha \alpha}$ is the Green's function of the isolated and biased $\alpha$-lead.
We wish to stress that a Green's function $\bg_{\alpha \alpha}$ 
with boundary conditions Eq. (\ref{kms}) automatically ensures the correct boundary 
conditions for the $\bcalG_{\alpha\rm C}(z,z')$ in Eq. 
(\ref{GalphaC}). Any other boundary 
conditions would not only lead to an unphysical transient behavior but 
also to different steady state results.\cite{sa.2004}
This is the case for, e.g., initially uncontacted
Hamiltonians in which the equilibrium chemical potential of the leads 
is replaced by the electrochemical potential, i.e., the sum of the 
chemical potential and the bias.

Taking into account Eq. (\ref{GalphaC}) the first term on the righthand side of Eq. (\ref{eomCC}) becomes
\be
\sum_{\alpha} \mathbf{H}_{\rm{C}\alpha} \bcalG_{\alpha \rm{C}} (z,z') =
\int d\bar{z} \,\bgS_{\rm{em}} (z,\bar{z}) \bcalG_{\rm{CC}} 
(\bar{z},z'),
\label{sig_em}
\ee
where we have introduced the \emph{embedding} self-energy
\be
\bgS_{\rm{em}} (z,z') = \sum_\alpha \bgS_{\rm{em},\alpha} (z,z') 
 = \sum_{\alpha}\mathbf{H}_{{\rm C}\alpha}\,
\mathbf{g}_{\alpha\alpha}(z,z') 
\mathbf{H}_{\alpha\rm C},
\label{embedding}
\ee
which accounts for the tunneling of electrons from the central region to the leads and vice versa. 
The embedding self-energies $\bgS_{\rm{em},\alpha}$ are independent of 
the electronic interactions and hence of $\bcalG_{\rm{CC}}$, 
and are therefore completely known once the lead Hamiltonians 
$\hH_\alpha$ of Eq. (\ref{eq:hamlead}) are specified.
Inserting (\ref{sig_em}) back to (\ref{eomCC}) then gives the equation of motion
\begin{equation}
\begin{split}
&\Bigl\{ i\de_z\mathbf{1} - \mathbf{H}_{\rm CC}(z) \Bigr\} 
\bcalG_{\rm CC}(z,z') \\
&=\delta(z,z')\mathbf{1} + \int d\bar{z}\,\left[ 
\bgS_{\rm CC}^{\rm MB}+
\bgS_{\rm em}\right](z,\bar{z})\,\bcalG_{\rm CC}(\bar{z},z').
\end{split}
\label{embeddedEOM}
\end{equation}
An adjoint equation can similarly be derived from Eq. (\ref{eqofmotion_adj}).
Equation (\ref{embeddedEOM}) is an exact equation for the Green's 
function $\bcalG_{\rm{CC}}$, for the class of Hamiltonians of Eq. (\ref{eq:ham}),
provided that an exact expression for $\bgS^{\rm{MB}}_{\rm CC}[\bcalG_{\rm{CC}}]$ 
as a functional of $\bcalG_{\rm{CC}}$ is inserted.
In practical implementations Eq. (\ref{embeddedEOM}) is converted to a set of
coupled real-time equations, known as the Kadanoff-Baym equations 
(see Appendix \ref{realtimekb}).
These equations are solved by means of time-propagation 
techniques.\cite{sdvl.2009b}
For the case of unperturbed systems the contributions of the integral 
in Eq. (\ref{embeddedEOM}) coming from the real-time branches
of the contour cancel and the integral needs only to be taken on the imaginary vertical track.
The equation for the Green's function then becomes equivalent to the one of the equilibrium finite-temperature formalism.
In a time-dependent situation the vertical track therefore accounts for initial correlations due 
to both many-body interactions, incorporated in 
$\bgS^{\rm{MB}}_{\rm CC}$, and contacts with the leads,
incorporated in $\bgS_{\rm{em}}$.
In our implementation (see Appendix \ref{realtimekb}) we always solve the {\em contacted} and {\em correlated} equation first on the
the imaginary track, before we propagate the Green's function in time in the presence of an external field.
However, to study initial correlations we are free to set the embedding and many-body self-energy
to zero before time-propagation, which is equivalent to neglect the vertical track of the contour.\cite{mssvl.2008}
This would correspond to starting with an equilibrium configuration that
describes an initially uncontacted and noninteracting central region. This class of initial configurations
is commonly used in quantum transport calculations, where both the interactions and the couplings
are considered to be switched on in the distant past. The assumption is then made that the system thermalizes
before the bias is switched on. Even when this assumption is fulfilled there are practical difficulties
to study transient phenomena, as one has to propagate the system until it has thermalized before a bias can be switched on.
It is therefore an advantage of our approach that thermalization assumptions are not necessary.\\
To solve the equation of motion Eq. (\ref{embeddedEOM}) we need to find an approximation for the many-body
self-energy $\bgS^{\rm{MB}}[\bcalG_{\rm{CC}}]$ as a functional of the Green's function $\bcalG_{\rm{CC}}$.
This approximation can be constructed using diagrammatic techniques based on Wick's theorem
familiar from equilibrium theory~\cite{FetterWalecka}
which can be straightforwardly be extended to the case of contour-ordered Green's functions.\cite{d.1984}
In our case the perturbative expansion is in powers of the two-body interaction and the unperturbed
system consists of the noninteracting, but contacted and biased system. We stress, however, that
eventually all our expressions are given in terms of fully dressed Green's functions leading to fully self-consistent
equations for the Green's function. This full self-consistency is essential to guarantee the satisfaction of the
charge conservation law, as is discussed in the next section.

\subsection{Charge conservation}
\label{chcons}

\begin{figure}[t]
\centering    
\includegraphics[width=0.45\textwidth]{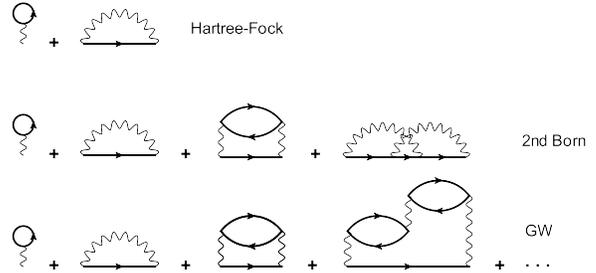}
\caption[]{Diagrammatic representation of the many-body 
approximations for $\bgS_{\rm CC}^{\rm MB}$.}
\label{fig3}
\end{figure}

The approximations for $\bgS^{\rm{MB}}_{\rm CC}[\bcalG_{\rm{CC}}]$ that 
we use in this work involve the Hartree-Fock, second Born and GW approximation,
which are discussed in detail in Refs. \onlinecite{dvl.2007,sdvl.2009,sdvl.2009b,dvl.2005} 
and are displayed pictorially in Fig. \ref{fig3}. 
These are all examples of so-called conserving approximations
for the self-energy, that guarantee satisfaction of fundamental conservation laws
such as charge conservation.
As shown by Baym~\cite{b.1962}, a self-energy approximation is conserving whenever it can be written as
the derivative of a functional $\Phi$, i.e.
\be
\bgS^{\rm{MB}}_{{\rm CC},rs} [\bcalG_{\rm{CC}}](z,z') = 
\frac{\delta \Phi [\bcalG_{\rm{CC}}]}{\delta \bcalG_{{\rm CC},sr} (z',z)}.
\label{eq:Phifunc}
\ee
This form of the self-energy is by itself not sufficient to guarantee that the conservation
laws are obeyed. A second condition is that the equations of motion for the Green's function
need to be solved fully self-consistently for this form of the self-energy 
(see, e.g., Ref. \onlinecite{sdvl.2009}).
For an open system, like our central region, charge conservation does 
not imply that the time derivative
of the number of particles $N_{\rm{C}}(t)$ is constant in time.
It rather implies that the time-derivative of $N_{\rm{C}}(t)$, also known as the
{\em displacement current}, is equal to the sum of
the currents that flow into the leads.
Below we give a proof in which the importance of the $\Phi$-derivability is clarified.
We start by writing
the number of particles $N_{\rm{C}}(t)$ as (see Eq.  (\ref{tddm}))
\be
N_{\rm C}(t) = -i \Tr_{\rm C}\left[ 
\bcalG_{\rm CC} (t_{-},t_{+})\right],
\ee
where the trace is taken over all one-particle indices in the central region.
Subtracting the equation of motion (\ref{embeddedEOM}) from its adjoint and
setting $z=t_{-}, z'=t_{+}$ then yields
\begin{equation}
\frac{d N_{\rm C}(t)}{d t} = -2{\rm Re}\,\Tr_{\rm C}\left[\int d\bar{z}
\,\bgS_{\rm{CC}}(t_-,\bar{z})\bcalG_{\rm{CC}}(\bar{z},t_+)\right],
\label{cc1}
\end{equation}
where $\bgS_{\rm{CC}}=\bgS_{\rm{CC}}^{\rm{MB}}+ \bgS_{\rm{em}}$.
By similar reasonings we can 
calculate the current $I_{\a}$ flowing across the interface between 
lead $\a$ and the central region. The total 
number of particles in lead $\a$ is $N_{\a}= -i \Tr_{\a}\left[ 
\bcalG_{\a\a} (t_{-},t_{+})\right]$, 
where the trace is taken over all one-particle indices in lead $\a$.
Projecting the equation of motion (\ref{eqofmotion}) on region
$\a \a$ yields
\beq
i \de_z \bcalG_{\a \a} (z,z') &=& \delta (z,z') \mathbf{1} + \mathbf{H}_{\a \a} (z) \bcalG_{\a \a} (z,z') \nonumber \\
&+& \mathbf{H}_{\a \rm C}(z) \bcalG_{{\rm C} \a} (z,z').
\label{eq:Gaa}
\eeq
Subtracting this equation
from its adjoint one finds
\beq
I_{\a}(t)&=&-\frac{d N_{\a}(t)}{d t}=2{\rm Re}\,
\Tr_{\a}\left[
\mathbf{H}_{\a\rm C}\bcalG_{{\rm C}\a}(t_{-},t_{+})\right]
\nonumber  \\
&=&2{\rm Re}\,\Tr_{\rm C}\left[
\bcalG_{{\rm C}\a}(t_{-},t_{+})\mathbf{H}_{\a\rm C}\right].
\eeq
Substituting in this expression the explicit solution 
(\ref{GalphaC}) for $\bcalG_{\a\rm C}$ as well as the solution for 
its adjoint $\bcalG_{{\rm C}\a}$ we can write the 
current $I_{\a}$ in terms of the embedding self-energy 
$\bgS_{\rm{em},\a}$ as
\be
I_{\a}(t)=2{\rm Re}\,\Tr_{\rm C}
\left[ \int d\bar{z}\,\bcalG_{\rm CC}(t_{-},\bar{z})\bgS_{{\rm 
em},\a}(\bar{z},t_{+}) \right].
\label{current}
\ee
Exploiting this result Eq. (\ref{cc1}) takes the form
\begin{equation}
\begin{split}
&\frac{d N_{\rm C}(t)}{d t} =I_{\rm L}+I_{\rm R} -\int d\bar{z}\,\Tr_{\rm C} \\
&\times\left[\bgS^{\rm MB}_{\rm{CC}}(t_-,\bar{z})\bcalG_{\rm{CC}}(\bar{z},t_+) - 
\bcalG_{\rm{CC}}(t_{-},\bar{z})\bgS^{\rm MB}_{\rm{CC}}(\bar{z},t_+) \right].
\end{split}
\label{cc3}
\end{equation}
Charge conservation implies that the integral in Eq. (\ref{cc3}) 
vanishes. This is a direct consequence of the 
invariance of the functional $\F$ under gauge 
transformations.
Indeed, changing the external potential by an arbitrary  purely time-dependent function
$\L_{r}(z)$ (with the boundary condition $\L_{r}(t_0)=\L_{r}(t_0-i\beta))$
changes the Green's function according to\cite{b.1962}
\begin{equation}
 \bcalG_{{\rm CC},rs}[\L](z,z') =  e^{i\L_{r}(z)}\bcalG_{{\rm CC},rs}(z,z')
e^{-i\L_{s}(z')},
\label{Ggauge}
\end{equation}
as can be checked directly from the equations of motion for the Green's function.
From its definition Eq. (\ref{eq:Phifunc}) it follows that the $\Phi$-functional
consists of closed diagrams in terms of the Green's function $ \bcalG_{\rm CC}$.
The phase factors of Eq. (\ref{Ggauge}) thus cancel each other at every vertex
and therefore $\Phi$ is independent of the functions $\L_{r}$.
This implies that
\beq
0&=&\sum_{q\in\rm C}\frac{\d\F}{\d \L_{q}(z)}
\nonumber \\
&=&
\sum_{qrs\in\rm C}\int d\bar{z}d\bar{z}'
\frac{\d\F}{\d \bcalG_{{\rm CC},sr}(\bar{z}',\bar{z})}
\frac{\d \bcalG_{{\rm CC},sr}(\bar{z}',\bar{z})}{\d \L_{q}(z)}
\nonumber \\
&=&
\sum_{qrs\in\rm C}\int d\bar{z}d\bar{z}'\,\bgS_{{\rm CC},rs}^{\rm 
MB}(\bar{z},\bar{z}')
\frac{\d \bcalG_{{\rm CC},sr}(\bar{z}',\bar{z})}{\d \L_{q}(z)},
\label{eq:dPhidL}
\eeq
where the sums run over all one-particle indices in the central region.
Here we explicitly used the $\Phi$-derivability condition of the self-energy
of Eq. (\ref{eq:Phifunc}).
If we now insert the derivative of the Green's function with respect to $\L_{r}$
from Eq. (\ref{Ggauge}) in Eq. (\ref{eq:dPhidL}) 
and evaluate the resulting expression in $z=t_{\pm}$ we obtain the integral 
in Eq. (\ref{cc3}). Therefore the last term in Eq. (\ref{cc3}) vanishes and the
time-derivative of the number of particles $N_{\rm C} (t)$ in the central
region is equal to the sum of the currents that flow into the leads.
We mention that in the long time limit the number 
of particles in region C is constant provided that the system 
attains a steady state. In this case $I_{\rm L}+I_{\rm R}=0$ and we 
recover the result of Ref. \onlinecite{tr.2008} as a special case.

\subsection{Equation for the time-dependent current}
\label{emwformula}

The time-dependent current in Eq. (\ref{current}) accounts for the initial 
many-body and embedding effects. In the absence of an external 
perturbation $I_{\a}(t)=0$ at any time. The exact vanishing of the current 
is guaranteed by the contribution of the vertical track in the 
integral. Discarding this contribution is equivalent to starting with an 
initially uncorrelated and uncontacted system in which case 
there will be some thermalization time during which  
charge fluctuations will give rise to nonzero transient currents.  \\
Equation (\ref{current}) involves an integral over the Keldysh contour. 
Using the extended Langreth theorem\cite{l.book,sa.2004,TDDFTbook} for the 
contour of Fig. \ref{fig2}
we can express $I_{\a}(t)$ in terms of real time and imaginary time integrals
\beq
I_{\alpha}(t)
=2\textnormal{Re}\,\Tr_{\rm C} \left[ 
\int_{t_0}^{t}d\bar{t}\,
\bcalG_{\rm CC}^{<}(t,\bar{t}) \bgS_{{\rm em},\a}^{A}(\bar{t},t)\right. 
\nonumber \\
+\int_{t_0}^{t}d\bar{t}\,
\bcalG_{\rm CC}^{R}(t,\bar{t}) \bgS_{{\rm em},\a}^{<}(\bar{t},t)
\nonumber \\
\left.
-i\int_{0}^{\b}d\t\,
\bcalG_{\rm CC}^{\rceil}(t,\t)\bgS_{{\rm em},\a}^{\lceil}(\t,t) \right],
\label{current2}
\eeq
where we refer to Appendix \ref{realtimekb} for the definition of the various superscripts.
Equation (\ref{current2}) provides a generalization of the Meir-Wingreen 
formula\cite{mw.1992} to the transient time-domain.
As anticipated the last term in Eq. (\ref{current2}) explicitly accounts for the effects
of initial correlations and initial-state dependence. If one
assumes that both dependences are washed out in the
long-time limit ($t\ra\inf$), then the last term in Eq. (\ref{current2})
vanishes and we can safely take the limit $t_0 \rightarrow -\infty$.
Furthermore, if in this limit the Green's function becomes a a function of the 
relative times only, i.e., $\bcalG_{\rm CC}(t,t') \ra \bcalG_{\rm CC} 
(t-t')$, we can Fourier transform with respect to the relative time 
to obtain the Green's function $\bcalG_{\rm CC}(\omega)$ and the 
self-energy $\bgS_{\rm{em}} (\omega)$  in frequency or energy space. 
This is typically the case for DC bias voltages where 
$\lim_{t\ra\inf}U_{\a}(t)=U_{\a}$. 
In terms of the Fourier transformed quantities Eq. (\ref{current2})  
reduces to the Meir-Wingreen formula\cite{mw.1992}
for the steady state current
\begin{equation}
I_{\alpha}^S =
-i\,\Tr_{\rm C} \int_{-\infty}^{\infty}\frac{d\omega}{2\pi} 
\bgG_{\alpha}(\omega)\Bigl\{   \bcalG_{\rm CC}^<(\omega) 
-2i\p f_{\alpha}(\omega) {\bf A}(\w) \Bigr\},
\label{mwcurrent}
\end{equation}
where
\be
\bgG_\alpha (\omega) = -2 \, \Im \, 
\{ \bgS_{\rm{em},\alpha}^R (\omega) \} ,
\ee
\be
{\bf A}(\w)=-\frac{1}{2\p i}[\bcalG_{\rm CC}^R(\omega)-\bcalG_{\rm 
CC}^A(\omega) ],
\label{matsf}
\ee
and where $f_\alpha$ is the Fermi distribution for lead $\alpha$ with 
electrochemical potential $\mu + U_\alpha$.
This expression has been used recently to perform steady state transport calculations
at GW level.\cite{tr.2007,tr.2008,t.2008} The present formalism allows for an extension
of this work to the time-dependent regime.

\subsection{Electron density in the leads}
\label{denslead}

In our investigations we are not only interested in calculating the density
in the central region, but are also interested in studying the densities in
the leads. In the following we will therefore derive an equation from which these
lead densities can be calculated. If we on the righthand side of Eq. (\ref{eq:Gaa})
insert the adjoint of Eq. (\ref{GalphaC}) we obtain the expression
\beq
i \de_z \bcalG_{\a \a} (z,z') &=& \delta (z,z') \mathbf{1} + \mathbf{H}_{\a \a} (z) \bcalG_{\a \a} (z,z')\nonumber \\
&+& \int d \bar{z} \bgS_{\rm{in},\alpha} (z,\bar{z}) \bg_{\a \a} 
(\bar{z},z') ,
\label{eq:Gaa_eom}
\eeq
where we defined the {\em inbedding} self-energy as
\be
\bgS_{\rm{in},\alpha} (z,z') = \mathbf{H}_{\a {\rm C}} 
\bcalG_{\rm{CC}} (z,z') \mathbf{H}_{{\rm C} \a}.
\ee
If we solve Eq. (\ref{eq:Gaa_eom}) in terms of $\bg_{\a \a}$ and take
the time arguments at $t_\pm$ we obtain
\beq
\bcalG_{\a \a} (t_-,t_+) &=& \bg_{\a \a} (t_-,t_+) + \nonumber \\
&+& \int d \bar{z} d \bar{\bar{z}} \bg_{\a \a} (t_-,\bar{z}) \bgS_{\rm{in},\a} 
(\bar{z},\bar{\bar{z}}) \bg_{\a \a} (\bar{\bar{z}},t_+).
\nonumber \\
\label{leaddens}
\eeq
We see from Eq. (\ref{tddm}) that
with this equation
we can obtain the spin occupation of orbital $i$ in lead $\alpha$ 
by taking $r=s=i\sigma \alpha$. The integral in Eq. (\ref{leaddens})
is taken along the Keldysh contour.
In practice we solve the Kadanoff-Baym equations for $\bcalG_{\rm{CC}}$
first. After this we construct the inbedding self-energy $\bgS_{\rm{in}}$
and calculate the lead density from Eq. (\ref{leaddens}) converted into
real time, using the conversion table of Ref. \onlinecite{TDDFTbook}.

\section{Numerical results}
\label{numres}

In this Section we specialize to central regions consisting of
quantum chains modelled using a tight-binding 
parametrization. We studied the case for which the chain extends from site 1 to site 4 
and is coupled to a left and right two-dimensional 
reservoirs with 9 transverse channels in the left and right leads, as illustrated in 
Fig. \ref{fig1}.
The parameters for the system are chosen as follows.
The longitudinal and transverse nearest neighbor hoppings in the leads are set to 
$T_\alpha^\lambda =T_\alpha^\tau = -2.0$, $\alpha={\rm L,R}$,
whereas the on-site energy $a^{\a}$ is set equal to the chemical 
potential, i.e., $a^\alpha=\mu$.
The leads are therefore half-filled. 
Precise definitions of these parameters can be found in Appendix \ref{emdse}.
The endsites of the central chain are coupled only to the terminal sites of the 
central row in both leads and the hopping parameters are 
$V_{1,5\rm L} = V_{4,5\rm R} = -0.5$ (see
Appendix \ref{emdse} for the labeling). 
The central chain has on-site energies $h_{ii}=0$ and hoppings
$h_{ij}=-1.0$ between neighboring sites $i$ and $j$.
The electron-electron interaction in the central region has the form $v_{ijkl} = 
v_{ij}\,\delta_{il}\delta_{jk}$ with 
\begin{equation}
v_{ij} =
\begin{cases}
v_{ii} & i=j\\
\frac{v_{ii}}{2|i-j|} & i\neq j 
\end{cases}
\end{equation}
and interaction strength $v_{ii} = 1.5$. For these parameters the 
equilibrium Hartree-Fock levels of the isolated chain lie 
at $\epsilon_1 = 0.39$, $\epsilon_2 = 1.32$, $\epsilon_3 = 3.19$, 
$\epsilon_4 = 4.46$. In all our simulations the chemical potential is 
fixed  between the highest occupied molecular orbital (HOMO) $\epsilon_2$ and 
the lowest unoccupied molecular orbital (LUMO) $\epsilon_3$ levels at 
$\mu = 2.26$ and the  inverse temperature $\beta$ is set to 
$\beta=90$ which corresponds to the zero temperature limit 
(i.e. results do not change anymore for higher values of $\beta$). 
In this work we will consider the case of a suddenly applied constant bias at an initial
time $t_0$, i.e. we take $U_\alpha (t)=U_\alpha$ for $t > t_0$ and  
$U_\alpha (t)=0$ for $t \leq t_0$.
Additionally, the bias voltage 
is applied symmetrically to the leads, i.e., $U_{L}=-U_{R}=U$, and 
the total potential drop is $2U$.

\subsection{Keldysh Green's functions in the double-time plane}
\label{kgreensect}

All physical quantities calculated in our work have been extracted 
from the different components of the Keldysh Green's function. 
Due to their importance we decided to present the behavior of the 
lesser Green's function $\bcalG^<$ as well as of the right Green's function 
$\bcalG^{\rceil}$ in the double-time plane for the Hartree-Fock 
approximation. The Green's functions corresponding to the 2B and GW
are qualitatively similar but show more strongly damped oscillations.
\begin{figure}[b]
\centering    
\includegraphics[width=0.4\textwidth]{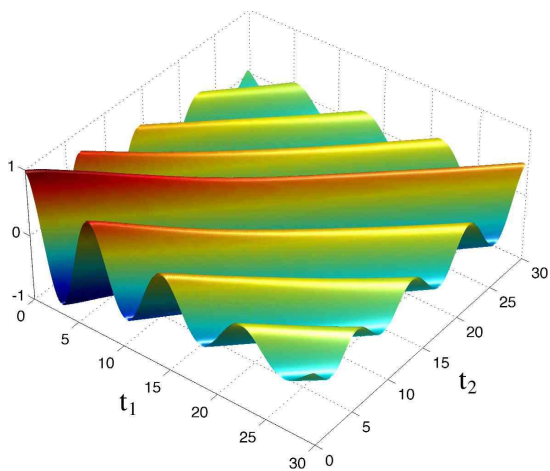}
\caption[]{The imaginary part of the lesser Green's 
function $\bcalG^<_{{\rm CC},HH}(t_1,t_2)$ of the central region in molecular orbital basis
corresponding to the HOMO level of the central chain.
Bias voltage $U=1.2$, HF approximation.}
\label{GlssHF}
\end{figure}
In Fig. \ref{GlssHF} we display the imaginary part of 
$\bcalG^<_{{\rm CC},HH}(t,t')$
in the basis of the initial Hartree-Fock molecular orbitals,  for an applied bias $U=1.2$. 
This matrix element corresponds to the HOMO level of the molecular chain.
The value of the Green's function on the time diagonal, i.e., 
$n_H (t) = \Im [\bcalG^<_{{\rm CC},HH}(t,t)]$
gives the level occupation number per spin.
We see that $n_H (t)$ decays from a value of $1.0$ at the initial time to a value of
$0.5$ at time $t=30$. An analysis of the LUMO level occupation $n_L (t)$ shows that
almost all the charge is transferred to this level.
The discharging of the
HOMO level and the charging of the LUMO level is also clearly observable
in the dipole moment as it causes a density oscillation in the system (see Section \ref{tddipmom}). 
When we move away from the time-diagonal we consider the time-propagation of
holes in the HOMO level. We observe a damped oscillation
the frequency of which corresponds to the removal energy of an electron
from the HOMO level,  leading to a distinct peak in the spectral function (see Section \ref{tdcurranddm}
below).
\begin{figure}[t]
\centering    
\includegraphics[width=0.4\textwidth]{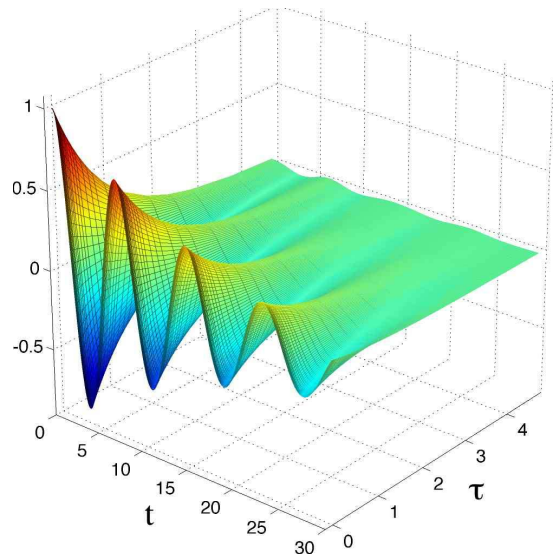}
\caption[]{The imaginary part of the mixed Green's 
function $\bcalG^{\rceil}_{{\rm CC},HH}(t,\tau)$ of the central region in 
molecular orbital basis. Bias voltage $U=1.2$, HF approximation.}
\label{G_rightHF}
\end{figure}

The imaginary part of $\bcalG^{\rceil}_{{\rm CC},HH}(t,\t)$  within the HF approximation
is displayed in Fig. \ref{G_rightHF} for real times
between $t=0$ and $t=30$ and imaginary times from $\tau=0$ to $\tau=5$. 
This mixed-time Green's function 
accounts for initial correlations as well as initial embedding effects
(within the HF approximation only the latter).
At $t=0$ we have the ground-state Matsubara Green's function and as
 the real time $t$ increases all elements of $\bcalG^{\rceil}_{\rm 
CC}(t,\t)$ approach zero independently of the value of $\t$. This 
behavior indicates that initial effects die out in the long-time limit
and that the decay rate is directly related to the time for reaching a steady 
state. A very similar behavior is found within the 2B and GW approximation
but with a stronger damping of the oscillations.

\subsection{Time-dependent current}
\label{tdcurranddm}

\begin{figure}[t]
\centering    
\includegraphics[width=0.45\textwidth]{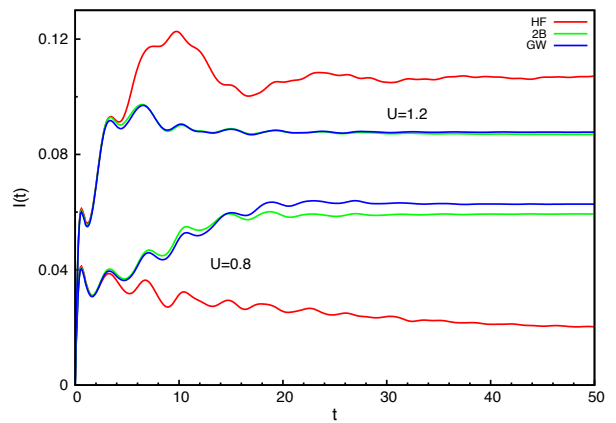}
\caption[]{Transient currents flowing into the right lead 
for the HF, 2B and GW approximations 
with the applied bias $U=0.8$ (three lowest curves) and $U=1.2$.}
\label{fig4}
\end{figure}

The time-dependent current at the right interface between the chain 
and the two-dimensional lead is shown in Fig. \ref{fig4} for the 
HF, 2B and GW approximations for two different values of the 
applied bias $U=0.8$ (weak) and $1.2$ (strong). 
The first remarkable feature is that the 2B and GW results are in 
excellent agreement {\em at all times} both in the weak and strong 
bias regime while the HF current deviates from the correlated results 
already after few time units. This result indicates 
that a chain of 4 atoms is already long enough for screening effects 
to play a crucial role. The 2B and GW approximations have in common 
the first three diagrams of the perturbative expansion of the many-body 
self-energy illustrated in Fig. \ref{fig3}. We thus conclude that 
the first order exchange diagram (Fock) with an interaction screened 
by an electron-hole propagator with a single polarization bubble
(with fully dressed Green's functions) contains the essential physics of 
the problem. We also wish to emphasize that the 2B approximation 
includes the so called second-order exchange diagram which is also 
quadratic in the interaction. This diagram is less relevant due to 
the restricted phase-space that two electrons in the chain have to 
scatter and exchange.

\begin{figure}[b]
\centering    
\includegraphics[width=0.45\textwidth]{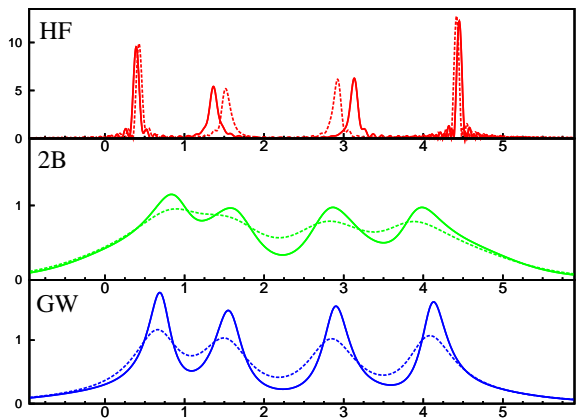}
\caption[]{Spectral functions $A(\omega)$ for HF (uppermost plot), 2B 
(middle plot) and GW (bottom plot) approximation with the
applied bias $U=0.8$ (solid line) and $U=1.2$ (dashed line).}
\label{fig6}
\end{figure}

We then turn our attention to the spectral function which is defined as
\be
A(T,\omega) = - \Im \Tr_{\rm C}  \int \frac{dt}{2 \pi} \, 
e^{i \omega  t} [ \bcalG_{\rm CC}^> - \bcalG_{\rm CC}^<] (T+ \frac{t}{2}, T -  \frac{t}{2}).
\ee
For values of $T$ after the transients have died out the spectral function becomes independent
of $T$. For such times we denote the spectral function by $A(\omega)$ 
and it is easy to show that $A(\w)=\Tr_{\rm C}[{\bf A}(\w)]$ where 
${\bf A}(\w)$ is defined in Eq. (\ref{matsf}).  This function displays peaks
that correspond to removal energies  (below the chemical potential) and 
electron addition energies (above the chemical potential).
The spectral functions of our system are displayed in Fig. \ref{fig6}.
At weak bias the HOMO-LUMO gap in the HF 
approximation is fairly the same as the equilibrium gap whereas the 
2B and GW gaps collapse causing both the HOMO and the LUMO to 
move in the bias window. As a consequence
the steady-state HF current is notably smaller than the 2B 
and GW currents. This effect has been previously observed by 
Thygesen\cite{t.2008} and is confirmed by our time-dependent 
simulations.

A new scenario does, however, emerge in the strong bias regime.
The HF HOMO and LUMO levels move into the bias window and
lift the steady-state current above the corresponding 2B and GW values. 
This can be explained by observing that the peaks of the 
HF spectral function $A(\omega)$ are very sharp compared to the rather 
broadened structures in the  2B and GW approximations, see Fig. 
\ref{fig6}. In the correlated case the HOMO and LUMO levels can be 
exploited only partially by the electrons to scatter from left to right 
and we thus observe a suppression of the current with respect to the HF case.
From a mathematical point of view the steady-state current is roughly 
proportional to the integral of $A(\omega)$ over the bias window which 
is larger in the HF approximation.

\begin{figure}[t]
\centering    
\includegraphics*[width=0.5\textwidth]{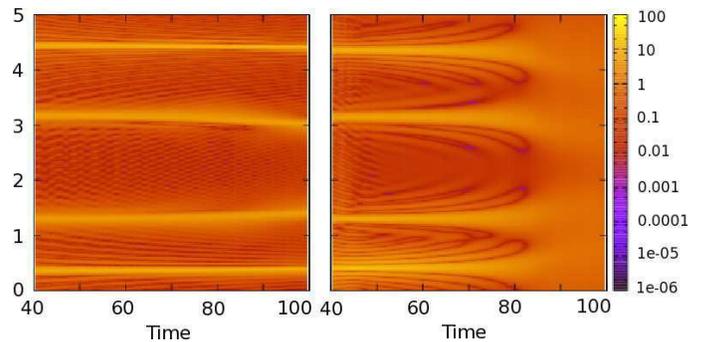}
\caption[]{Real-time evolution of the spectral function $A(T,\omega)$ for the HF (left panel) and the 2B approximation (right panel)
for an applied bias of $U=1.2$. On the horizontal axis the time $T$ and the vertical axis the frequency $\omega$. }
\label{realtimespectrumHF}
\end{figure}

The time-evolution of the spectral function $A(T,\omega)$ as a function of $T$ 
 is illustrated in Fig. \ref{realtimespectrumHF} for the case of the HF and the 2B  approximation. 
For these results, the ground state system was propagated without bias 
up to $T=40$ after which a bias was suddenly turned on. 
The HF peaks remain rather sharp during the entire 
evolution and the HOMO-LUMO levels come nearer to each other at 
a constant speed. On the contrary, the broadening of 
the 2B peaks remains small during the initial transient regime 
(up to $T=70$) to then increase dramatically. This behavior indicates that 
there is a critical charging time after which an enhanced
renormalization of quasiparticle states takes place causing 
a substantial reshaping of the equilibrium spectral function. 

\begin{figure}[t]
\centering    
\includegraphics[width=0.50\textwidth]{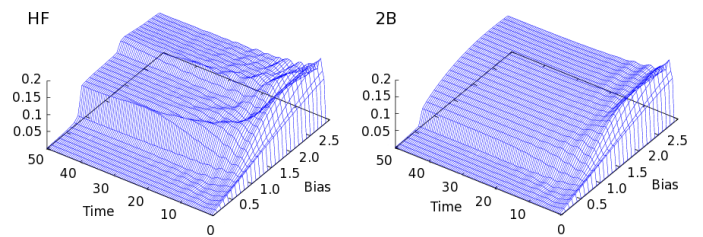}
\caption[]{Transient right current $I_{\rm R}(U,t)$ as a function of applied bias 
voltage and time in the HF (left panel) and 2B (right panel) 
approximations.}
\label{fig9}
\end{figure}
\begin{figure}[t]
\centering    
\includegraphics[width=0.5\textwidth]{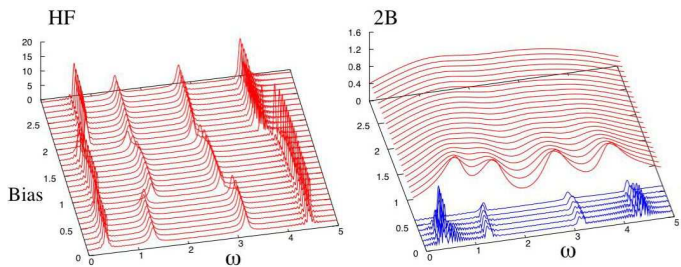}
\caption[]{Spectral function $A(\omega)$ for the HF (left panel) and 2B (right panel) approximation, as a function of the bias voltage.
For the 2B approximation the spectral functions for bias voltages until $U=0.6$ were divided by a factor 30 (blue lines in the figure)}
\label{fig10}
\end{figure}

The time-dependent current at the right interface as a function 
of applied voltage and time is shown in Fig. \ref{fig9} for 
the HF and 2B approximation. 
The figures nicely illustrate how steady state results are obtained
from time-dependent calculations:
after the transients have died out
we see the formation of the characteristic I-V curves familiar from steady
state transport calculations.
In the HF approximation one clearly observes the  typical
staircase structure with steps that correspond to an applied voltage 
that includes one more resonance in the bias window.  These steps appear 
at bias voltages $U=0.9$ and  $U=1.8$.
This result is corroborated by the left panel of Fig. \ref{fig10} in which we display 
the bias-dependent spectral function for the HF approximation. Here we
see a sudden shift in the spectral peaks at these voltages.
The HF results thus bear a close resemblance to the standard 
non-interacting results, the main difference being that
the HF position of the levels gets renormalized by the applied bias.

We now turn our attention to the 2B approximation in the right panel of Fig.\ref{fig9}. 
We notice a clear step at bias voltage of $U=0.7$ but the
broadening of the level peaks due to quasiparticle collisions completely
smears out the second step and
the current increases smoothly as a function of the applied voltage.
This is again corroborated in the right panel of Fig.\ref{fig10} where we
observe a sudden broadening of the spectral function at a bias of $U=0.7$.
To make this effect clearly visible in the figure we divided the spectral functions for biases
up to $U=0.6$ by a factor of $30$.
We further notice that for the 2B approximation there is a faster gap closing as a function
of the bias voltage as compared to the HF approximation. 
Very similar results are obtained within the GW approximation.
We can therefore conclude that electronic correlations beyond Hartree-Fock level have a 
major impact on both transient and steady-state currents.

\subsection{Time-dependent dipole moment}
\label{tddipmom}

\begin{figure}[b]
\centering    
\includegraphics[width=0.45\textwidth]{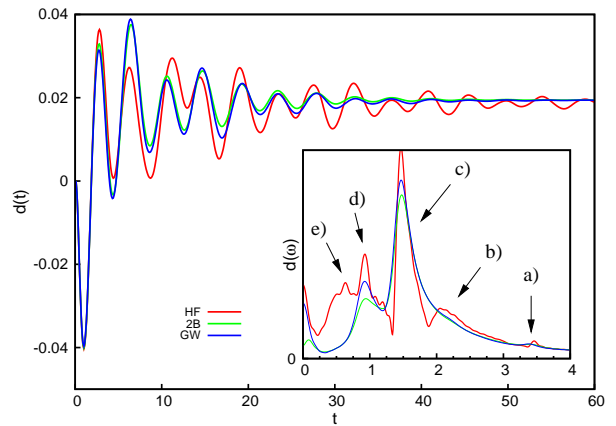}
\caption[]{Dipole moment of the central region as a function of time 
for bias $U=1.2$. 
The inset shows the Fourier transform of the dipole moment.}
\label{fig11}
\end{figure}

To study how the charge redistribute along the chain after a bias 
voltage is switched on we calculated the time-dependent 
dipole moment 
\begin{equation}
d(t) = \sum_{i=1}^{4}x_in_i(t)
\end{equation}
where the $x_{i}$ are the coordinates of the sites of the chain 
(with a lattice spacing of one) with  origin between sites 2 and 3. 
As observed in Section \ref{kgreensect}  
the chain remains fairly charge neutral during the entire time 
evolution. However, a charge rearrangement occurs as can be seen from 
Fig. \ref{fig11}. At $U=1.2$ both the HOMO and the LUMO are inside the 
bias window, the lowest level remains below and the highest 
level above. Electrons in the initially populated HOMO then move to the 
empty LUMO and get only partially reflected back. This 
generates damped oscillations with the HOMO-LUMO gap as the main 
frequency, a non-vanishing steady value for the LUMO population
and a partially filled HOMO. Due to the 
different (odd/even) approximate spatial symmetry of the HOMO/LUMO levels 
a net dipole moment develops.

As we pointed out in a recent Letter,\cite{mssvl.2008} 
the oscillations in the transient current reflect the electronic transitions 
between the {\em ground state} levels of the central region and the 
electrochemical potentials of the left and right leads. However, the oscillations
are visible in all observable quantities through the oscillations of 
the Green's function discussed in Section \ref{kgreensect}. 
Detailed information on the 
electronic level structure of the chain can be extracted from the Fourier 
transform of $d(t)$, see inset in Fig. \ref{fig11}. One clearly recognize 
the presence of sharp peaks superimposed to a broad continuum. 
The peaks occur at energies corresponding to 
electronic transitions from lead states at the left/right electrochemical 
potential to chain eigenstates or to intrachain transitions.
We will denote a transition energy between leads ${\rm L}$ and ${\rm R}$ and chain eigenstate $i$ by
$\Delta \epsilon_{{\rm L}i}$ and $\Delta \epsilon_{i{\rm R}}$. Similarly we will
denote a transition energy between states in the central region as $\Delta \epsilon_{ij}$.
In the inset of Fig. \ref{fig11} the main peak structures are labeled from the highest to the lowest transition
energies with letters (a) to (e) and we will use these labels to denote the various transitions
discussed below. The possible transition energies can be determined  form the position
of the peaks in the spectral functions and the lead levels.
As expected the dominant peak occurs at the intrachain transition energy 
$\Delta\epsilon_{23}\approx 1.5$ (c). This roughly corresponds 
to the average of the equilibrium and nonequilibrium gaps and, 
therefore, must be traced back to charge fluctuations  
between the HOMO and LUMO. 
The other observable transition energies are 
$\Delta\epsilon_{{\rm L}2}\approx 2.0$ (b), $\Delta\epsilon_{{\rm L}3}\approx 0.5$ (e) and 
$\Delta\epsilon_{{\rm L}4}\approx 1.0$ (d) from the left 
lead and
$\Delta\epsilon_{1{\rm R}}\approx 0.65$ (e), $\Delta\epsilon_{2{\rm R}}\approx 0.4$ (e), 
$\Delta\epsilon_{3{\rm R}}\approx 2.0$ (b) and 
$\Delta\epsilon_{4{\rm R}}\approx 3.4$ (a) from the right lead. Some of the peaks
with transition energies close to each other 
($\Delta\epsilon_{{\rm L}2}$ \& $\Delta\epsilon_{3{\rm R}}$ (b) and 
$\Delta\epsilon_{{\rm L}3}$ 
\& $\Delta\epsilon_{1{\rm R}}$ \& $\Delta\epsilon_{2{\rm R}}$(e)) are merged together and 
broadened. The broadening is not only due to  embedding and 
many-body effects but also to the dynamical renormalization of the 
position of the energy levels.
Further information can be extracted from the peak intensities. 
The peak of the $\Delta\epsilon_{{\rm L}4}$ (d) transition
is very strong due to the sharpness of that particular 
resonance, see Fig. \ref{fig6}, and its initial low population.
On the contrary, the transition $\Delta\epsilon_{{\rm L}1}$ 
from the left lead to the highly 
populated level $\epsilon_{1}$ is extremely weak due 
to the Pauli blockade and not visible.
Correlation effects beyond Hartree-Fock theory causes a fast
damping of all sofar discussed transitions. Only the transitions 
$\Delta\epsilon_{{\rm L}4}$ (d) and  $\Delta\epsilon_{23}$ (c) are
visible in the Fourier spectrum of the 2B and GW approximation.

\subsection{Time dependent screened interaction $W$}
\label{tdw}

\begin{figure}[h]
\centering    
\includegraphics[width=0.4\textwidth]{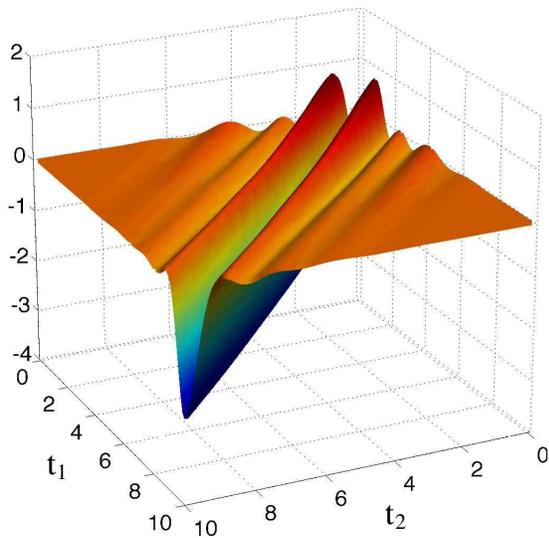}
\caption[]{Imaginary part of the trace of the screened interaction 
$W^<(t_1,t_2)$ in the GW approximation. }
\label{Screenedinteraction}
\end{figure}

In Fig. \ref{Screenedinteraction} we show the trace of the lesser component of the time-dependent 
screened interaction of the GW approximation in the double-time plane.
This interaction is defined as $W =v + v\,P\,W$ 
where $P$ is the full polarization bubble\cite{sdvl.2009} (with dressed Green's functions) of 
the connected and correlated system, and gives information on the 
strength and efficiency of the dynamical screening of 
the repulsive interactions. The good agreement between 
the 2B and GW approximations implies that the 
dominant contribution to the screening comes from the
first bubble diagram, that is $W^< \approx vP^<v$. 
From Fig. \ref{Screenedinteraction} we see that the trace of the 
imaginary part of $W^{<}(t,t)$ is about 3. Considering that the trace 
of the instantaneous bare interaction $v$ is 6 we conclude that the screening 
diagrams reduce the magnitude of the repulsion by a factor of 2.
Another interesting feature of the screened interaction is that it 
decays rather fast when the separation of the time arguments 
increases. From Fig. \ref{Screenedinteraction} we see that after a 
time $t\approx 7$ the retarded interaction is negligibly small.
It is worth noting that such a time scale is much smaller than the 
typical time scales to reach a steady state, see Fig. \ref{fig4}.

\subsection{Time-dependent Friedel oscillations in the leads}
\label{tdfriedel}

We implemented the method described in Section \ref{denslead} and 
based on the {\em inbedding} technique to investigate the electron 
dynamics in the leads. This study is of special importance since it 
challenges one of the main assumption in quantum transport 
calculations, i.e., that the leads remain in thermal equilibrium 
during the entire evolution.

\begin{figure}[b]
\centering    
\includegraphics[width=0.45\textwidth]{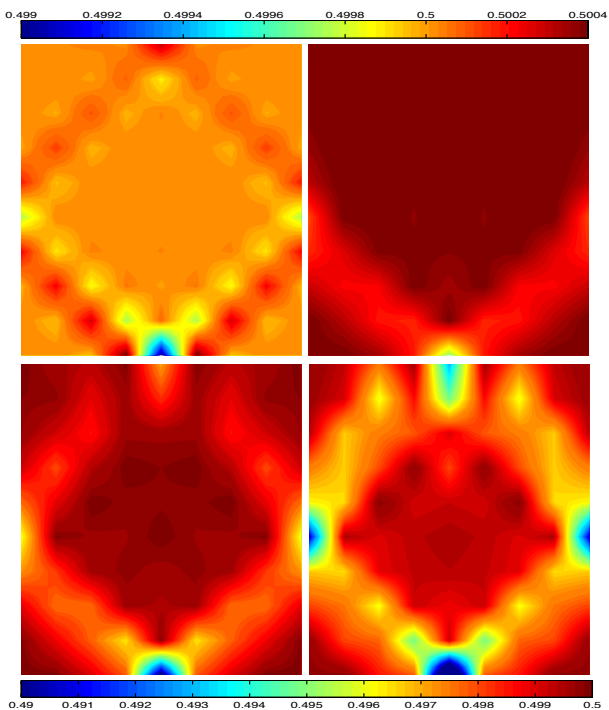}
\caption[]{Snapshots of the density in left lead for 
HF approximation after the bias $U=1.2$ switch-on.
On the horizontal axes the transverse dimension of the 
lead (9 rows wide, with the site connected to the chain in the center)
and 10 layers deep.
Upper panel left: Initial density, Upper panel right: 
density at time $t=1.7$, Lower left panel: density at time $t=3.6$,
Lower right panel: density at time $t=10$.
The upper colorbar refers to the initial density in the upper left panel.
The lower colorbar refers to the remaining pictures.}
\label{leaddensity}
\end{figure}

In Fig. \ref{leaddensity} we show the evolution of the density in 
the two-dimensional 9-row wide leads 
(see Fig. \ref{fig1}) after the sudden switch-on of a bias voltage.
We display snapshots of the lead densities at times
$t=0,1.7,3.6$ and $10$ where up to
10 layers deep into the leads (where to improve the visibility we 
interpolated the density between the sites).
Since the atomic wire is connected to the central site it
acts as an impurity and we see density oscillations in the leads
following diamond-like pattern. These present Friedel oscillations
that propagate along preferred directions. 

The preferred directions in the density pattern can be understood from 
linear response theory. Given a square lattice with nearest neighbor 
hopping $T=T^{\l}=T^{\t}$ the retarded density response function in Fourier space reads
\beq
\chi({\bf q},\w)&=&\int\frac{d{\bf k}}{(2\p)^{2}}
\frac{f(\e_{{\bf k}})-f(\e_{{\bf k}+{\bf q}})}
{\w-\e_{{\bf k}}+\e_{{\bf k}+{\bf q}}+i\eta}
\nonumber \\ 
&=&
2\int\frac{d{\bf k}}{(2\p)^{2}}
\frac{f(\e_{{\bf k}})(\e_{{\bf k}}-\e_{{\bf k}+{\bf q}})}
{(\w+i\eta)^{2}-(\e_{{\bf k}}-\e_{{\bf k}+{\bf q}})^{2}},
\label{eq:densresponse}
\eeq
where $\e_{{\bf k}}=2T(\cos k_{x}+\cos k_{y})$ is the energy 
dispersion and the integral is done over the first Brillouin zone and 
$f$ is the Fermi distribution function.
At half filling the Fermi energy is zero and the Fermi surface is a square with vertices in 
$(0,\pm \p)$ and $(\pm\p,0)$. The dominant contribution to the 
integral comes from the values of ${\bf k}$ close to such vertices 
where the density of states has van Hove singularities. 
The response function $\chi({\bf q}=\a{\bf Q},\w=0)$, with ${\bf 
Q}=(\p,\p)$ the nesting vector, is discontinuous for $\a=1$. Indeed, 
for every occupied ${\bf k}$ there exists an $\a<1$ such that 
$\e_{{\bf k}+{\bf q}}=\e_{{\bf k}}<0$ and the integrand diverges at 
zero frequency. On the other hand for $\a>1$ the vector ${\bf k}+{\bf 
q}$ corresponds to an unoccupied state with energy 
$\e_{{\bf k}+{\bf q}}>0$ and due to the presence of the Fermi function the 
integrand of Eq.(\ref{eq:densresponse}) is well behaved even for 
$\w=0$. The discontinuity at  ${\bf Q}=(\p,\p)$ 
is analogous to the discontinuity at $2k_{F}$ in the electron gas and 
leads to the Friedel oscillations with diamond symmetry
observed in Fig. \ref{leaddensity}. By adding reciprocal lattice vectors
we find that there are four equivalent directions for these Friedel oscillations given
by the vectors ${\bf Q}=\pm (\pi, \pm\pi)$. Each of these vectors gives in real space
rise to a density change of the form $\delta n(\mathbf{r}) \sim e^{{i \mathbf{Q}} \cdot \mathbf{r}}$. 
Therefore a single impurity in a 2D lattice
induces a cross-shaped density pattern. Due to the fact that
in our case the lattice ends at the central chain, we only observe two arms of
this cross.

The results of Fig. \ref{leaddensity} also allows for testing 
the assumption of thermal equilibrium in the leads. 
The equilibrium density [Top-left panel] is essentially the same as 
its equilibrium bulk value 
at 0.5. After the switching 
of the bias a density corrugation with the shape of a diamond starts 
to propagate deep into the lead. The largest deviation from the bulk 
density value occurs at the corners of the diamond and is about  
$2\%$ at the junction while it reduces to about $1\%$ after 10 layers.
We also verified that the discrepancy is about 3 times larger for leads 
with only three  transverse channels. We conclude that the change in 
the lead density goes like the inverse of the cross section. 
Our results suggests that for a mean 
field description of 2D leads with 9 transverse channels it is enough to 
include few atomic layers for an accurate self-consistent time-dependent 
calculations of the Hartree potential.

\section{Conclusions}
\label{conc}

We proposed a time-dependent many-body approach based on the 
real-time propagation of the KB equations to tackle quantum transport 
problems of correlated electrons. We proved the continuity equation 
for any $\F$-derivable self-energy, a fundamental 
property in non-equilibrium conditions, and generalize the 
Meir-Wingreen formula to account for initial correlations and initial 
embedding effects. This
requires an extension of the Keldysh contour with the thermal
segment $(t_{0},t_{0}-i\b)$ and the consideration of mixed-time Green's functions 
having one real and one imaginary time argument. The Keldysh Green's 
function in the device region $\bcalG_{\rm{CC}}$ is typically used to calculate 
currents and densities in the device. In this work we also developed an 
exact {\em inbedding} scheme to extract from $\bcalG_{\rm{CC}}$ the TD 
density in the leads.

The theoretical framework and the implementation scheme were tested for
one-dimensional wires connected to two-dimensional leads using 
different approximations for the many-body self-energy. We found that 
already for 4-sites wires screening effects play a crucial role. 
The 2B and GW approximations are in excellent agreement {\em at all 
times} for moderate interaction strength (of the same order of 
magnitude of the hopping integrals) while the HF approximation tends 
to deviate from the GW and 2B results after very short times. These 
differences were
related to the sharp peaks of the HF spectral function as compared to 
the rather broad structures observed in 2B and GW. Our numerical 
results indicate that the largest part of the correlation effects are 
well described by the first bubble diagram of the self-energy, common 
to both the 2B and GW approximation. The screened interaction was 
explicitely calculated in the GW approximation showing that the 
screening reduces the interaction strenght by a factor of 2 and that 
retardation effects are absent after a time-scale much shorter than 
the typical transient time-scale.
The electron dynamics obtained using a correlated self-energy differ 
from the HF dynamics in many respects:
1) At moderate bias the HOMO-LUMO gap closes while in the HF 
approximation it remains fairly constant;
2) The HOMO and LUMO resonances are rather sharp during the transient 
time to then suddenly broaden when approaching the steady state. 
This indicates the occurrence of an enhanced renormalization of 
quasiparticle states. The HF widths instead remain unaltered.
3) The transient time in the correlated case is much shorter than in 
HF, see Fig. \ref{fig11}.

The transient behavior of time-dependent quantities like the current and 
dipole moment exhibit oscillations of characteristic frequencies that
reflect the underlying level-structure of the system.
Calculating the ultrafast response of the device to an external 
driving field thus constitutes an alternative method 
to gain insight into the quasi-particle positions and life-times
out of equilibrium. We performed a discrete Fourier 
analysis of the TD dipole-moment in the transient regime and related 
the characteristic frequencies to transitions either between 
different levels of the wire 
or between the levels of the wire and the electrochemical 
potential of the leads. The hight of the peaks in the Fourier 
transform can be interpreted as the amount of density which oscillate 
between the levels of a given transition. In all approximations we 
found that the density mainly sloshes between the HOMO and the LUMO.

One of the main assumption in quantum transport calculations is that 
the leads remain in thermal equilibrium and therefore that the 
bulk density is not affected by the presence of the junction. 
To investigate this assumption we considered two-dimensional leads 
thus going beyond the so called wide-band-limit approximation. 
By virtue of an exact {\em inbedding} technique we calculated the 
lead density both in and out equilibrium.
In the proximity of the 
junction the density exhibits Friedel-like oscillations 
whose period depend on the value of the Fermi momentum along the given 
direction.

In conclusion the real-time-propagation of the KB equations for 
open and inhomogeneous systems provide a very powerful tool to study 
the electron dynamics of a typical quantum transport set-up. 
In this work we considered only DC biases. However, more complicated driving 
fields like AC biases or pumping fields can be dealt with at the same 
computational cost and the results will be the subject of a future 
publication. Besides currents and densities the MBPT 
framework also allows for calculating higher order correlators.
It is our intention to use the KB equations to study 
shot-noise in quantum junctions using different levels of 
approximation for the Green's function.

\appendix
\section{The embedded Kadanoff-Baym equations}
\label{realtimekb}

To apply Eq. (\ref{embeddedEOM}) in practice we need to transform it to real-time equations
that we solve by time-propagation.
This can be done in Eq. (\ref{embeddedEOM})
by considering time-arguments of the Green's function and self-energy on different branches of the contour.
We therefore have to define these components first.
Let us therefore consider a function on the Keldysh contour of the general form
\beq
F(z,z') &=& F^\delta (z)\delta (z,z') \nonumber \\
 &+& \theta (z,z') F^> (z,z') + \theta (z',z) F^< (z,z') ,
\eeq
where $\theta (z,z')$
is a contour Heaviside function,\cite{d.1984} i.e. $\theta (z,z')=1$
for $z$ later than $z'$ on the contour and zero otherwise,
and $\delta (z,z')=\partial_z \theta (z,z')$ is the contour delta function.
By restricting the variables $z$ and $z'$ on different branches of the contour
we can define the various
components of $F$ as
\beq
F^\lessgtr (t,t') &=& F(t_{\mp},t_{\pm}'), \\
F^{\rceil} (t,\tau) &=& F(t_{\pm},t_0-i\tau), \\
F^{\lceil} (\tau,t) &=& F(t_0-i\tau,t_{\pm}) ,\\
F^M (\tau - \tau') &=& -i F(t_0-i\tau, t_0-i\tau'),
\eeq
and 
\be
F^{R/A} (t,t') = F^\delta (t) \delta (t-t') \mp \theta (\pm t \mp t') [F^>(t,t')-F^< (t,t')].
\ee
For the Green's function there is no singular contribution, i.e., $\bcalG^\delta=0$, but the
self-energy has a singular contribution of Hartree-Fock form, i.e.,
$\bgS^\delta = \bgS^{\rm HF} [\bcalG]$.\cite{d.1984} 
With these definitions we can now convert Eq. (\ref{embeddedEOM}) to
equations for the separate components. This is conveniently done using the conversion table
in Ref. \onlinecite{TDDFTbook}.
We then obtain the following set of equations
\beq
i\de_t \bcalG^\lessgtr(t ,t') &=& \mathbf{H}_{\rm{CC}}(t) \bcalG^\lessgtr(t,t')
+\left[\bgS^{R} \cdot \bcalG^\lessgtr\right] (t,t')
\nonumber\\ 
&+& \left[\bgS^\lessgtr \cdot \bcalG^A \right] (t,t') + \left[\bgS^\rceil \star \bcalG^\lceil \right](t,t'), 
\nonumber \\
\label{eq:kb1} \\
-i\de_{t'} \bcalG^\lessgtr(t, t') &=&  \bcalG^\lessgtr(t,t') \mathbf{H}_{\rm{CC}}(t')
+\left[\bcalG^{R} \cdot \bgS^\lessgtr\right] (t,t')
\nonumber\\ 
&+& \left[\bcalG^\lessgtr \cdot \bgS^A \right] (t,t') + \left[\bcalG^\rceil \star \bgS^\lceil \right](t,t'), 
\nonumber \\
\label{eq:kb2} \\
i\de_t\bcalG^\rceil(t,\tau) &=& \mathbf{H}_{\rm CC}(t) \bcalG^\rceil(t,\tau)
+\left[\bgS^R \cdot \bcalG^\rceil\right] (t, \tau) \nonumber \\
&+& \left[\bgS^\rceil \star \bcalG^\mathrm{M} \right](t,\tau),
\label{eq:kb3} \\
-i\de_t\bcalG^\lceil(\tau,t) &=&  \bcalG^\lceil(\tau,t) \mathbf{H}_{\rm CC}(t)
+\left[\bcalG^\lceil \cdot \bgS^A \right] (\tau,t) \nonumber \\
&+& \left[\bcalG^M \star \bgS^\lceil \right](\tau,t),
\label{eq:kb4} \\
- \de_{\tau} \bcalG^M (\tau-\tau') &=& \mathbf{1} \delta (\tau-\tau') + \mathbf{H}_{\rm{CC}} \bcalG^M (\tau-\tau') \nonumber \\
&+& i \left[\bgS^M \star \bcalG^\mathrm{M} \right](\tau-\tau'),
\label{eq:Matsubara}
\eeq 
which are commonly known as the Kadanoff-Baym equations.
The symbols $\cdot$ and $\star$ are a shorthand notation for the real-time and imaginary-time convolutions
\begin{equation}
\begin{split}
&\left[ a\cdot b \right](t,t') = \int_0^{\infty} 
a(t,\bar{t})b(\bar{t},t') d\bar{t},\\
&\left[ a\star b \right](t,t') = -i \int_0^{\beta} a(t,\tau)b(\tau,t') d\tau.
\end{split}
\end{equation}
In practice we first solve Eq. (\ref{eq:Matsubara}) which describes the initial equilibrium Green's function.
This equation is decoupled from the other two, since $\bgS^M$ depends on $\bcalG^M$ only.
The initial conditions for the other Green's functions $G^\lessgtr$ and $G^{\rceil \lceil}$ are then
determined by $\bcalG^M$ as follows
\beq
\bcalG^> (0,0) &=& i \bcalG^M (0^+), \\
\bcalG^< (0,0) &=& i \bcalG^M (0^-), \\
\bcalG^\rceil (0,\tau) &=& i \bcalG^M (-\tau), \\
\bcalG^\lceil (\tau,0) &=& i \bcalG^M (\tau).
\eeq
With these initial conditions the Eqs.(\ref{eq:kb1})-(\ref{eq:kb4}) 
can be solved using a time-stepping algorithm.\cite{sdvl.2009b}

\section{Embedding self-energy}
\label{emdse}

From Eq. (\ref{embedding}) and Eq. (\ref{HalphaC}) we see that the embedding self-energy has the form
\be
\bgS_{{\rm em},\alpha,kl} (z,z') = \sum_{ij} V_{k,i\alpha} \bg_{\a 
\a,ij} (z,z') V_{j\alpha,l},
\label{embedding2}
\ee
where $k$ and $l$ label orbitals in the central region.
As can be seen from this equation, the calculation of the embedding self-energy requires
the determination of $\bg_{\a\a}$.
Since for the isolated lead $\alpha$ the time-dependent field is simply a gauge,
$\bg_{\a\a}$ is of the form
\be
\bg_{\a \a} (z,z') = \bg_{\a \a}^0 (z,z') \exp \left( -i \int_{z'}^z 
d\bar{z} \, U_{\a} (\bar{z}) \right),
\ee
where $\bg_{\a \a}^0$ is the Green's function for the unbiased lead, and the
integral in the exponent is a contour integral. 
The Green's function $\bg^0_{\a \a}$ has the form
\be
\bg_{\a \a}^0 (z,z') = \theta (z,z') \bg^{0,>}_{\a \a}(z,z') +  
\theta (z',z) \bg^{0,<}_{\a \a}(z,z'). 
\ee
It therefore remains to
specify $\bg^{0,\lessgtr}_{\a \a}$.
In the following we will for convenience separate out the spin part 
from the Green's function and write $\bg^0_{\a\a,i\sigma,j\sigma'}=\delta_{\sigma \sigma'}\bg^0_{\a \a,ij}$.
We will now give give an explicit expression for $\bg^0_{\a \a,ij}$
for the case of two-dimensional leads. The case of three dimensions can be treated
similarly. We consider a lead Hamiltonian of a tight-binding form,
that is separable in the longitudinal ($x$) and the transverse ($y$) directions. Therefore
the indices in the one-particle matrix $h_{ij}^\alpha$ of Eq. (\ref{eq:hamlead}) denote sites $i=(x, y), j=(x', y')$
where $x$ and $y$ are integers running from zero to $N_x^\alpha$ and $N_y^\alpha$. At the end of the derivation we
take the limit $N_x^\alpha \rightarrow \infty$.
The Hamiltonian matrix for the leads is then of the form
\be
h^\alpha_{ij} (t) = \delta_{x x'} \tau^\alpha_{y y'} + \delta_{y y'} 
\lambda^\alpha_{x x'} + a^\alpha \delta_{ij},
\label{eq:hlead}
\ee
where $\lambda$ and $\tau$ are matrices that represent longitudinal and transverse chains and $a^\alpha$ is an on-site energy.
Hence
\beq
\bg_{\a \a, ij}^{0,\lessgtr} (z,z') = \sum_{p} U^\alpha_{ip}  \, 
\bg^{0,\lessgtr}_{\a \a,p} (z,z')  U_{p j}^{\alpha \dagger} ,
\label{eq:gsitebasis}
\eeq
where $p=(p_x,p_y)$ is a two-dimensional index spanning the same one-particle space.
The matrix $U^\alpha=D^{\tau \alpha} \otimes D^{\lambda \alpha}$ is a 
direct product of the unitary matrices $D^{\tau \alpha}$ and $D^{\lambda \alpha}$
that diagonalize the matrices $\tau^\alpha$ and $\lambda^\alpha$ in Eq. (\ref{eq:hlead}) 
The functions $\bg^{0,\lessgtr}_{\a \a,p}$ have the explicit form
\beq
\bg_{\a \a,p}^{0,<} (z,z') &=& i f(\epsilon_{p \alpha}) e^{ -i 
\int_{z'}^z d\bar{z} \, (\epsilon_{p \alpha}-\m) } ,\\
\bg_{\a \a,p}^{0,>} (z,z') &=& i (f(\epsilon_{p \alpha}) -1) e^{ 
-i \int_{z'}^z d\bar{z} \, (\epsilon_{p \alpha}- \mu) },
\eeq
with $f(\epsilon)=1/(e^{\beta (\epsilon-\m)}+1)$ the Fermi distribution function.
In these expressions $\epsilon_{p\alpha} = \epsilon_{p_y \alpha}^\tau 
+ \epsilon_{p_x \alpha}^\lambda$,
where $\epsilon_{p_y \alpha}^\tau$ and $\epsilon_{p_x 
\alpha}^\lambda$ are the eigenvalues of matrices $\t^\alpha$ 
and $\l^\alpha$.
\begin{figure}[t]
\centering    
\includegraphics[width=0.48\textwidth]{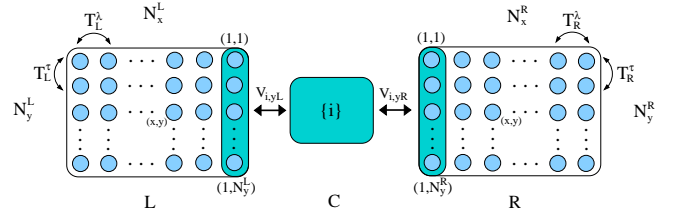}
\caption[]{Tight-binding system for finite 2D leads connected to scattering central region.}
\label{fig15}
\end{figure}
In the case the matrices $\tau^\alpha$ and $\lambda^\alpha$ represent tight-binding chains with nearest neigbour hoppings
$T_\alpha^\tau$  and $T_\alpha^{\lambda}$ and zero on-site energy, we have
\beq
D^{\lambda \alpha}_{x p_x} &=& \sqrt{\frac{2}{N_x^\alpha+1}} \sin 
(\frac{\pi x p_x}{N_x^\alpha + 1}), \\
\epsilon_{p_x \alpha}^\lambda &=& 2 T_\alpha^\lambda \cos (\frac{\pi 
p_x}{N_x^\alpha+1}),
\eeq
and similarly for the transverse transformation matrix $D^{\tau \alpha}_{y p_y}$ and energy $\epsilon_{p_y \alpha}^\tau$.
If we insert these expressions in Eq. (\ref{eq:gsitebasis}) and take the limit $N_x \rightarrow \infty$ such that we
can replace summation over $p_x$ by an integration over the angular variable
$\phi=\pi p_x / (N_x^\alpha +1)$, then we obtain
\beq
\bg_{\a \a,ij}^{0,<} (z,z') &=&  \frac{4i }{N_y^\alpha+1}  \sum_{p_y=1}^{N_y^\alpha} \sin (\frac{\pi y p_y }{N_y^\alpha+1}) \sin (\frac{\pi y' p_y }{N_y^\alpha+1})\nonumber \\
&\times& \frac{1}{\pi} \int_0^\pi d\phi
\sin (x \phi) \sin (x' \phi) \nonumber \\
&\times& 
f(\epsilon_{p \alpha}) e^{ -i \int_{z'}^z d\bar{z} \, 
(\epsilon_{p \alpha}- \mu) } ,
\label{gless}
\eeq
where now
\be
\epsilon_{p\alpha} = a^\alpha + 2 T_\alpha^\tau \cos (\frac{\pi p_y}{N_y^\alpha+1}) +  2 T_\alpha^\lambda \cos \phi.
\ee
The expression for $\bg_{\a \a,ij}^{0,>}$ is obtained from Eq. (\ref{gless}) by simply replacing the Fermi function $f$
by $f-1$. Let us now turn to the embedding self-energy.
In this work we consider the case that we only have hopping elements $V_{i,k\alpha}$  between central sites $i$  and the
first tranverse layer of the leads, which are labeled by elements $k=(1,y)$ where $y=1 \ldots N_y^\alpha$. 
However, the entire formalism can extended to more general cases.
This means that we take 
\beq
V_{i,k\alpha}= \left\{ \begin{array}{cc} V_{i,y\alpha} & \mbox{if $k=(1,y)$} \\ 0 & \mbox{otherwise}  \end{array} \right..
\eeq
In that case in Eq. (\ref{gless}) only the contribution with $x=x'=1$ survives. 
Then the product of the sine functions can be written in terms of the eigenenergies of
the isolated leads as
\beq
 \bgS_{{\rm em},\alpha,kl}^< (z,z') &=& \sum_{y,y',p_y=1}^{N_y^\alpha} 
\frac{4i V_{k,y\alpha} V_{y'\alpha,l} }{N_y^\alpha+1}   \nonumber \\
&\times& \sin (\frac{\pi y p_y }{N_y^\alpha+1}) \sin (\frac{\pi y' p_y }{N_y^\alpha+1})\nonumber \\
&\times& \frac{1}{\pi} \int_{E_{p_y \alpha}^-}^{E_{p_y \alpha}^+} \frac{d\epsilon}{2 |T_\alpha^\lambda|} 
\sqrt{ 1 - \left(\frac{E_{p_y \alpha}}{2 T_\alpha^\lambda}\right)^2 }
\nonumber \\
&\times& 
f(\epsilon) e^{ -i \int_{z'}^z d\bar{z} \, (\epsilon- \mu) } ,
\label{embedding3}
\eeq
where we defined $E_{p_y \alpha} = \epsilon - a^\alpha - \epsilon_{p_y \alpha}^\tau$
and $E_{p_y \alpha}^\pm= a^\alpha + \epsilon_{p_y \alpha}^\tau \pm 2 |T_\alpha^\lambda|$.
The expression for $\bgS_{{\rm em},\a}^{>}$ is obtained from Eq. (\ref{embedding3}) by simply replacing the Fermi function $f$
by $f-1$. In the case that there is no transverse coupling, i.e., 
$T_\alpha^\tau=0$, the integral is independent of
$p_y$ and the sum over $p_y$ can be performed to yield $\delta_{y 
y'}$. Then the 2D self-energy becomes
a sum of self-energies over separate 1D leads.

\end{document}